\documentclass[preprint, double-spaced]{revtex4}
\usepackage[colorlinks=true,linktocpage=true,linkcolor=blue,citecolor=blue]{hyperref}
\usepackage{amsmath, amsfonts, amsthm, amssymb}
\usepackage{graphicx}
\usepackage{amsmath,bbm}
\usepackage{amssymb,bm}
\usepackage{array}
\usepackage{float}
\usepackage{chngpage}
\usepackage{boxedminipage}
\usepackage{enumerate}
\usepackage{multirow}

\begin{document}
\title{Universal Freezeout Condition for Charged Hadrons in a Hybrid Approach}
\author{O. S. K. Chaturvedi$^1$}
\author{P. K. Srivastava$^{2,}$\footnote{prasu111@gmail.com}}
\author{Arpit Singh$^1$}
\author{B. K. Singh$^{1,}$\footnote{bksingh@bhu.ac.in}}
\affiliation{$^1$Department of Physics, Institute of Science, Banaras Hindu University, Varanasi-221005, INDIA}
\affiliation{$^2$Department of Physics, Indian Institute of Technology Ropar, Rupnagar- 140001, INDIA}
\begin{abstract}
Hadronic freezeout during the evolution of the medium created in heavy-ion collisions is an important phenomena. It is quite useful to find a universal freezeout condition for each and every nuclear collisions. In this article, we have constructed a hybrid model to calculate the ratio of transverse energy to total mean multiplicity $E_{T} /N_{ch}$, since this ratio can possibly act as a freezeout condition in heavy-ion collision experiments. Present hybrid model blends two approaches : Tsallis statistics and wounded quark approach. Recently, Tsallis statistics has been reliably used to obtain the transverse momentum distribution of charged hadrons produced in relativistic ion collisions. On the other side it has been shown that the pseudorapidity distribution of charged hadrons can be calculated satisfactorily using the wounded quark model (WQM). We have used this hybrid model to calculate the transverse energy density distributions, $dE_{T}/d\eta$ at midrapidity using charged particle pseudorapidity distributions, $dN_{ch}/d\eta$ and mean transverse momentum $\langle p_{T} \rangle$ in various type of nuclear collisions. We found that present hybrid model satisfactorily explains the experimental data whether other models fail to reproduce the data at central and at peripheral collisions simultaneously. Finally, ratio of transverse energy to total mean multiplicity, $E_{T} /N_{ch}$ has been computed within hybrid model and compared with the available experimental data at RHIC and LHC energies. We observed no explicit dependence of $E_{T} /N_{ch}$ on energy as well as centrality and thus it can definitely act as a freezeout criteria.

\end{abstract}

\maketitle 
\section{Introduction}
\noindent
The hadronic matter undergoes a phase transition from hadrons to a deconfined phase characterized by quarks and gluons, at extreme temperatures and/or energy density. This deconfined phase which can be created by colliding nuclei at ultrarelativistic energies ~\cite{Singh_Reports:0993,SinghIJMA:1992,Dremin:2001,Gribov_83,Levin_90}, is known as quark gluon palsma (QGP). The static and dynamic properties of the fireball created in these heavy ion collisions can be understood by investigating the global observables like pseudorapidity density ($dN_{ch}/d\eta$), transverse energy density ($dE_{T}/d\eta$), transverse momentum ($p_{T}$) spectra, transverse energy per unit multiplicity, $E_{T}/N_{ch}$ [$\equiv(dE_{T}/d\eta)/(dn_{ch}/d\eta)$], etc.~\cite{Cleymans:2007uk,Mishra:2013dwa,Sahoo:2014aca,skt2,Adams:2004cb,Adam:2016thv,Adler:2004zn}. The analysis of these observables with respect to various control parameters e.g. centrality, collision energy, transverse momentum, pseudorapidity etc., can provide a better understanding of multiparticle production mechanism and about this exotic phase. For example, transverse energy density can provide a better understanding of the collision reaction dynamics. Further multiplicity and the shape of pseudorapidity distribution of charged hadrons can provide a hint for the QGP formation as we know that multiplicity is directly related with the initial entropy and its evolution ~\cite{Sahoo:2014aca,Prorok:2004af,Kharzeev:2000ph}.

In nucleus-nucleus collisions, multiparticle production shows a basic characteristic that most of the particles are created in longitudinal direction with a large longitudinal momentum ($p_{L}$) and a small transverse momentum ($p_{T}$). In comparison to the longitudinal direction, small number of particles are created in transverse direction with large $p_{T}$ and small $p_{L}$. Therefore one can study the produced charged hadron distribution with respect to either $p_{T}$ or $y$($\eta$). In both cases the distribution shows an exponential behaviour which is either $exp(-p_{T}/T)$ or $exp(-E/T)$ in $p_{T}$ and $y$-space, respectively. However the basic difference is in the value of parameter $T$ which is dependent on available energy and ranges in GeV in $y-$ or $\eta -$distribution while it is independent of energy and ranges in MeV in $p_{T}$-distribution. This implies that these two distribution arises due to different physical mechanism. $p_{T}$-space is ``thermal-like'' and $p_{L}$-space is sensitive to available energy and the multiplicity of charged secondaries ~\cite{Wilk_2009,Wilk_2012,Rybczynski_2014,liuAuAu2014, pPbAzmi2013,wilk201405,Khandai:2014}.

The pseudorapidity distribution data is well confronted by two-component wounded nucleon model like Glauber model or two-component wounded quark model ~\cite{Shyam_1:1989,Shyam:1985,Singh_3:1986,miller_2007,Deng_2011,Levin_2010,Armesto_2005,Sarkisyan_2010, Ashwini_2:2013,Ashwini:2013,Chaturvedi:2016,Chaturvedi_2017}. These models basically explore the idea of law of equipartition of available energy between participant nuclei or quarks and the number of charged secondaries produced in $p$-$p$ collision. However, in recent years it has been shown that wounded quark model is more appropriate for particle production in comparison to wounded nucleon model ~\cite{Bozek_2012,Bozek_2016,Bialas_2008,Loizides_2016,Barej_arX_2017,Bozek_2017}. On the other hand, $p_{T}$ distribution shows the deviation from the exact thermal distributions. Practitioners in this field have come up with the idea of dynamical effects like flow on the $p_{T}$-distribution. Recently another idea of non-equilibrium effects on $p_{T}$-distribution has come into light. In this approach, it has been suggested that the deviation from the exact thermal distribution to a power law distribution is due to the intrinsic, nonstatistical fluctuations. These deviations can be properly included by introducing Tsallis statistical approach in multiparticle production process in $p_{T}$-space. Thus the Tsallis distribution which originates from Tsallis statistics became a good candidate to study the transverse momentum distributions in nucleus-nucleus collisions at (ultra-)relativistic energies ~\cite{liuAuAu2014,pPbAzmi2013,wilk201405,Khandai:2014,huapp,deflow2014}. Further using the average transverse momentum derived from Tsallis statistics enable us to study the transverse energy density distribution which again makes our understanding about the particle production mechanism better.

In connection to the particle production mechanism, the ratio of transverse energy density and the pseudorapidity density of produced charged particles is an important observable in high energy heavy-ion collisions. This ratio is a measure of the mean transverse energy per particle and reveals about the mechanism of hadronic freeze-out. Its collision energy and centrality dependence is exactly like the chemical freeze-out temperature up to highest relativistic energies. A lot of study has already been done to understand the various scenarios of chemical freezeout i.e., single freezeout, multiple freezeouts etc.~\cite{Cleymans:2007uk,Cleymans:2005xv,chatterjee:2017:1,chatterjee:2017:2}. If the ratio $E_{T}/N_{ch}$ remain almost constant in our study then it will possibly support the idea of single freezeout surface for all the non-strange charged hadrons produced in various collisions at different energies. However, recently it has been discussed that the ratio $E_{T}/N_{ch}$ may be affected by the collectivity and non-equilibrium phenomena or the effect of gluon saturation, which is expected at higher collision energies specially at LHC energies ~\cite{Sahoo:2014aca,Prorok:2004af,Kharzeev:2000ph,Cleymans:2005xv,Cleymans:1998fq,Cleymans:2008it}. Thus it is much needed to calculate this $E_{T}/N_{ch}$ ratio and study its behaviour with respect to collision energy and centrality.

The main aim of this article is to construct a hybrid model to calculate the mean transverse momentum and study the transverse energy density distribution along with a ratio $E_{T}/N_{ch}$ which can possibly act as a freezeout criteria. In the present hybrid model we amalgate our newly proposed version of wounded quark model (WQM) to calculate  pseudorapidity distributions and Tsallis approach to fit transverse momentum distributions of charged particles and to calculate average transverse momentum at midrapidity. In our recent papers ~\cite{Ashwini_2:2013,Ashwini:2013,Chaturvedi:2016,Chaturvedi_2017}  we have shown that our WQM properly describes the total mean multiplicities and pseudorapidity distributions for variety of collision species e.g., symmetric, asymmetric and deformed systems etc., over a wide range of collision energies and with collision centrality from peripheral to most central events. Further it has been shown in recent literatures that Tsallis power-law distribution satisfies the transverse momentum distribution quite well for various type of collisions. Thus we blend these two appropriate approach to study transverse energy density distribution and the freezeout criteria, as mean transverse energy per particle $E_{T}/N_{ch}$.

As we already know from earlier literatures that the wounded quark models are good in predicting pseudorapidity distributions in comparison to wounded nucleon models and Tsallis distribution is also good for fitting $p_{T}$-distributions. Thus it is important here to clearly state new and interesting features of our present work as follows : (1) Our version of WQM is quite different from other versions of wounded quark model as it uses two component approach instead of other versions which says that the multiplicity simply scales with one component, i.e., mean number of wounded quarks. We have chosen two component WQM since we have shown in one of our recent paper~\cite{Chaturvedi_2017} that in peripheral collisions one component WQM is not sufficient and we need a small fraction of second component which depends on the mean number of quark-quark collisions. (2) In our work we have used a simple form of Tsallis distribution instead of complicated form to fit the $p_{T}$ distributions so that we can able to present the physical significance of the fitting parameters specially in nucleus-nucleus collisions. As most of the literatures, explained the physical significance of parameters in Tsallis distribution only for $p-p$ collisions. (3) We have presented a unified, consistent and a comprehensive hybrid model which satisfies almost all the data from collision experiments of various species and at different energies. (4) The most important part of the present article is that we have used this unified and consistent hybrid model to calculate the $E_{T}/N_{ch}$ and shown its variation quite rigorously with colliding species, energy and centrality which is scarcely studied in earlier literatures. Further we have obtained an important result in present paper that $E_{T}/N_{ch}=constant$ can act as a universal freezeout criteria for particle production in the collisions of large as well as small colliding systems in a wide range of energies.

The rest of the paper is organized as follows: In section II, we start with the modified Tsallis parametrization for transverse momentum distribution of charged hadrons and details of the method to calculate average transverse momentum. Further we provide a brief description of the formulation of wounded quark model for calculating the pseudorapidity density with respect to centrality for variety of collision species at RHIC and LHC energies. Furthermore, we will provide the expression to calculate the transverse energy density of the charged hadrons. In Section III we will discuss the results consisting pseudorapidty distribution, average transverse momentum, transverse energy density and the freezeout criteria as $E_{T}/N_{ch}$. At last we will summarize our present work.

\section{Model Formalism}
\subsection{Modified Tsallis distribution}
Tsallis distribution provides the useful information of the transverse momentum distributions of produced particles in hadronic as well as nuclear collisions. The low-$p_{T}$ part of the spectra is controlled by the processes in which the momentum transfer is small and the coupling constant is large. Therefore, this region is dominated by non-perturbative QCD physics. However in high-$p_{T}$ region, the coupling constant is small and it is usually considered as a perturbative QCD regime where hard scattering between a parton of one hadron and a parton of other hadron produces the charged hadrons ~\cite{huapp, Khandai:2014, li2015ad, wilk201405, liuAuAu2014, pPbAzmi2013, daupip2006,cleymans, azmiJPG2014,  liAuAu2013, maciej, wongprd, wong2012, wongarxiv2014}. Tsallis statistics provide us a tool to develop a nonextensive formula which works in the whole $p_{T}$ range and is given as follows:
\begin{eqnarray}
E\frac{d^3N}{dp^3}&=&\frac{1}{2\pi p_T} \frac{d^2N}{dydp_T} \nonumber\\ 
&=& \frac{dN}{dy} \frac{(n-1)(n-2)}{2\pi nC[nC+m(n-2)]}(1+\frac{m_T-m}{nC})^{-n},\nonumber\\
\label{exptsallis}
\end{eqnarray}
where $m_T=\sqrt{p_T^2+m^2}$ is the transverse mass and $m$ is the mass of the particle.  $\frac{dN}{dy}$, $n$ and $C$ are fitting parameters.
In literature several people have used the thermodynamic consistent form of Tsallis distribution ~\cite{cleymans, azmiJPG2014, pPbAzmi2013, liAuAu2013, maciej, liufh20147} given below:
\begin{eqnarray}
E\frac{d^3N}{dp^3} = gV\frac{m_T \cosh y}{(2\pi)^3} [1+(q-1)\frac{m_T\cosh y-\mu}{T}]^{\frac{q}{1-q}}, \label{tsallisB}
\end{eqnarray}
where $g$ is the degeneracy of the particle, $V$ is the volume, $y$ is the rapidity, $\mu$ is the chemical potential, $T$ is the temperature and $q$ is a parameter. The form of Eq. 2 in the mid-rapidity $y=0$ region is reduced as
\begin{equation}
E\frac{d^3N}{dp^3} = gV\frac{m_T}{(2\pi)^3} [1+(q-1)\frac{m_T}{T}]^{\frac{q}{1-q}}. \label{tsallisBR}
\end{equation}

In Ref. ~\cite{wong2012}, Wong {\it et al.} proposed a new form of the Tsallis distribution function to take into account the rapidity cut as
\begin{equation}
(E\frac{d^3N}{dp^3})_{|\eta|<a}=\int_{-a}^{a}d\eta \frac{dy}{d\eta}(\frac{d^3N}{dp^3}). \label{tsalliswong}
\end{equation}
where
\begin{eqnarray}
\frac{dy}{d\eta}(\eta, p_T)=\sqrt{1-\frac{m^2}{m^2_T\cosh^2 y}}, 
\end{eqnarray}
with the rapidity variable defined as,
$$y=\frac{1}{2}\ln \Big [\frac{\sqrt{p_T^2 \cosh^2 \eta + m^2}+p_T\sinh \eta}{\sqrt{p_T^2 \cosh^2 \eta + m^2}-p_T\sinh \eta}\Big],$$
and the $\frac{d^3N}{dp^3}$ is given as,
\begin{equation}
\frac{d^3N}{dp^3}=C\frac{dN}{dy}(1+\frac{E_T}{nT})^{-n},\quad 
E_T=m_T-m, 
\end{equation}
where $C\frac{dN}{dy}$ is assumed to be a constant parameter.

Now, one can obtained a simplified form as given in Ref.  ~\cite{huapp},
\begin{equation}
(E\frac{d^3N}{dp^3})_{|\eta|<a} = A(1+\frac{E_T}{nT})^{-n}, \label{tsallisus}
\end{equation}
where $A$, $n$ and $T$ are the fitting parameters. In the present calculation we have used the Eq. (7) as a fitting function to fit the experimental data for variety of collision species.

As given in Ref. ~\cite{wilk201405}, one has to slightly modify the above equation to fit the particle spectra in $Pb$-$Pb$ collision at $\sqrt{s_{NN}}$ = 2.76 TeV. In the modified parametrization the number of parameters increased by one and fitted the experimental $p_{T}$ distribution data very accurately. Here, we have used the similar modified form having four free parameters and can be expressed as follows :

 \begin{equation}
 (E\frac{d^3N}{dp^3})_{|\eta|<a}=A \frac{e^{-\frac{b}{T}\arctan(E_T/b)}}{[1+(\frac{E_T}{b})^4]^{c}}. \label{neweq}
 \end{equation}
One can see the asymptotic behaviour of the above equation, when $\frac{E_T}{b}<<1$,
\begin{equation}
(E\frac{d^3N}{dp^3})_{|\eta|<a}\propto e^{-\frac{E_T}{T}},
\end{equation} 
and, when $\frac{E_T}{b}>>1$, 
\begin{equation}
(E\frac{d^3N}{dp^3})_{|\eta|<a}\propto p_T^{-4c}.
\end{equation} 

\subsection{Average Transverse Momentum of Charged Hadrons}
The average transverse momentum, $\left \langle p_{T} \right \rangle$ at midrapidity can be calulated using the invarient yield as ~\cite{Ohsawa_2010}  
\begin{equation}
\left \langle p_{T} \right \rangle_{y=0}=\int dp_{T}p_{T}\left ( d^{2}N_{ch}/dydp_{T} \right )/\int dp_{T}\left (d^{2}N_{ch}/dydp_{T} \right) 
\end{equation}
The variable $d^{2}N_{ch}/dydp_{T}$ can be suitably compared through Eq. 1, to obtained the form in terms of free fitting parameters as used in Eq. 7. For each colliding system we got the fitting parameters value as given in Table I and II and using this we integrate over the available $p_{T}$ range at RHIC and LHC energy with proper normalization factor to the get the value of $\left \langle p_{T} \right \rangle_{y=0}$.

\begin{table*}
\begin{center}
 \caption{The fitting parameters and the corresponding $\chi^2$/ndf values for charged pions in different collision systems at different collision energy with Tsallis distribution as in Eq. 7. }
 \begin{tabular}{c c c c c c c}
    \hline
    System &Particle &Centrality &A &T(GeV) &n & $\chi^{2}/ndf$ \\
    \hline
    \multirow{5}{6cm}{Au+Au $\sqrt{s_{NN}}$ = 200 GeV} & \multirow{2}{1.2cm}{$\pi^{+}$} &0-12 &1210 &0.1334 &10.12 &316.9/26 \\
                                                    &                            &20-40 &531.2 &0.1289 &9.46 &313.1/26 \\ 
                                                    & \multirow{2}{1.2cm}{$\pi^{-}$} &0-12 &1200 &0.1332 &10.05 &325.2/26 \\
                                                    &                            &20-40 &566 &0.1267 &9.35 &324.6/26 \\
    \hline
    \multirow{7}{6cm}{Au+Au $\sqrt{s_{NN}}$ = 62.4 GeV} & \multirow{3}{1.2cm}{$\pi^{+}$} &0-10 &537.3 &0.1734 &17.18 &16.65/20 \\
                                                      &                           &10-20 &383.12 &0.1701 &16.25 &19.9/19 \\
                                                      &                           &20-40 &225.5 &0.1664 &15.37 &22.37/20 \\ 
                                                      & \multirow{3}{1.2cm}{$\pi^{-}$} &0-10 &549.9 &0.1731 &17.25 &13.71/20 \\
                                                      &                           &10-20 &388.2 &0.1706 &16.43 &16.36/19 \\
                                                      &                           &20-40 &224.7 &0.1681 &15.66 &14.21/20 \\
    \hline
    \multirow{7}{6cm}{d+Au $\sqrt{s_{NN}}$ = 200 GeV} & \multirow{3}{1.2cm}{$\pi^{+}$} &0-20 &14.4 &0.1715 &10.15 &20.17/21 \\
                                                   &                            &20-40 &11.29 &0.1675 &10.13 &33.41/21 \\
                                                   &                            &40-60 &8.406 &0.1624 &10.10 &30.75/21 \\ 
                                                   & \multirow{3}{1.2cm}{$\pi^{-}$} &0-20 &13.64 &0.1744 &10.33 &28.98/21 \\
                                                   &                            &20-40 &10.69 &0.1702 &10.31 &18.87/21 \\
                                                   &                            &40-60 &8.001 &0.1649 &10.29 &32.56/21 \\
    \hline
    \multirow{9}{6cm}{Cu+Cu $\sqrt{s_{NN}}$ = 200 GeV} & \multirow{4}{1.2cm}{$\pi^{+}$} &0-10 &475 &0.1275 &9.778 &8.099/8 \\
                                                    &                            &10-20 &340 &0.1224 &9.30 &8.066/8 \\
                                                    &                            &20-40 &215 &0.12 &9.16 &8.1/8 \\
                                                    &                            &40-60 &200 &0.1089 &9.14 &8.402/8 \\ 
                                                    & \multirow{4}{1.2cm}{$\pi^{-}$} &0-10 &498 &0.1209 &9.35 &8.041/8 \\
                                                    &                            &10-20 &395 &0.1176 &9.19 &8.037/8 \\
                                                    &                            &20-40 &325 &0.1113 &8.99 &8.058/8 \\
                                                    &                            &40-60 &250 &0.1012 &8.73 &8.146/8 \\ 
    \hline
    \multirow{5}{6cm}{p+Pb $\sqrt{s_{NN}}$ = 5.02 TeV} & \multirow{5}{1.2cm}{charged $\pi$} &0-5 &81.59 &0.1752 &7.336 &232.2/55 \\
                                                    &                            &5-10 &67.53 &0.1717 &7.206 &231.8/55 \\
                                                    &                            &10-20 &59.28 &0.1672 &7.073 &218.3/55 \\
                                                    &                            &20-40 &48.25 &0.1616 &6.991 &202.9/55 \\
                                                    &                            &40-60 &37.11 &0.1524 &6.853 &158.6/55 \\
    \hline
  \end{tabular}
  
\end{center}
\end{table*}

\begin{table*}
\begin{center}
 \caption{The fitting parameters and the corresponding $\chi^2$/ndf values for charged pions in $Pb$-$Pb$ system at LHC energy with Tsallis distribution as in Eq. 8.}
 \begin{tabular}{c c c c c c c c}
    \hline
    System &Particle &Centrality &A &T(GeV) &b &c & $\chi^{2}/ndf$ \\
    \hline
    \multirow{5}{6cm}{Pb+Pb $\sqrt{s_{NN}}$ = 2.76 TeV} & \multirow{2}{1.2cm}{charged $\pi$} &0-5 &2050 &0.252 &2.195 &0.886 &173.1/59 \\
                                                    &                            &5-10 &1780 &0.245 &2.025 &0.901 &310.4/59 \\
                                                    &                            &10-20 &1438 &0.239 &1.894 &0.916 &520.3/59 \\
                                                    &                            &20-40 &815.2 &0.236 &1.795 &0.925 &802/59 \\
                                                    &                            &40-60 &291.8 &0.234 &1.699 &0.939 &781.3/59 \\
    \hline
  \end{tabular}
  
\end{center}
\end{table*}

\subsection{Pseudorapidity from WQM}
In the recent articles ~\cite{Ashwini_2:2013,Ashwini:2013,Chaturvedi:2016,Chaturvedi_2017}, we have formulated a new version of wounded quark model which provides satisfactorily results regarding charged hadron production in hadronic as well as nuclear collisions. Here, we have used the two-component wounded quark model to obtain the pseudo-rapidity distribution in nucleus-nucleus collisions. The simple assumption behind the two component WQM is that the hard component, which basically arises due to multiple parton interactions ~\cite{Trainor_ar}, scales with the number of quark-quark collisions (i.e. $N_{q}^{AB}\nu_{q}^{AB}$) and soft component scales with the number of participating quarks (i.e.$N_{q}^{AB}$). So, we used the expression for $\left(\frac{dn_{ch}}{d\eta}\right)^{AB}_{\eta=0}$ in $A$-$B$ collisions as parametrized in terms of $p$-$p$ rapidity density ~\cite{Ashwini_2:2013,Ashwini:2013} is given below, 
\begin{equation}
\left(\frac{dn_{ch}}{d\eta}\right)^{AB}_{\eta=0}=\left(\frac{dn_{ch}}{d\eta}\right)^{pp}_{\eta=0}\left[\left(1-x\right)N_{q}^{AB}+ x N_{q}^{AB}\nu_{q}^{AB}\right],
\end{equation}
Here, $x$ quantifies the relative contributions of two components arising from hard and soft processes.
Taking the assumption of additive quark model~\cite{BialasPRD:1982,Anisovich:1984,lipkin}, we have calculated the $\nu_{q}^{AB}$ in following manner, 
\begin{equation}
\nu_{q}^{AB}=\nu_{qA}\nu_{qB}=\frac{A\sigma_{qN}^{in}}{{\sigma_{qA}^{in}}}.\frac{B\sigma_{qN}^{in}}{{\sigma_{qB}^{in}}}.
\end{equation}
where, $\nu_{qA}$ is the mean number of inelastic quark collisions in nucleus A, and $\sigma_{qN}^{in}$ is the quark-nucleon inelastic cross section, and $\sigma_{qA}^{in}$ is the quark-nucleus inelastic cross-section.
Now, the mean number of participating quarks $N^{AB}_{q}$ is defined as,
\begin{equation}
N^{AB}_{q}=\frac{1}{2}\left[\frac{N_{B}\sigma_{qA}^{in}}{{\sigma_{AB}^{in}}}+\frac{N_{A}\sigma_{qB}^{in}}{{\sigma_{AB}^{in}}}\right],
\end{equation}
where $\sigma_{AB}^{in}$ is the inelastic cross-section for $A$-$B$ collisions. Further, to calculate $\sigma_{AB}$ we take the help of optical model as discussed in Refs. ~\cite{fernbach,hoang} and can be expressed in the following manner:
\begin{equation}
\sigma_{AB}^{in}= \pi r^{2}\left[A^{1/3}+B^{1/3}-\frac{c}{A^{1/3}+B^{1/3}}\right]^2.
\end{equation}
Here, the constant $c$ is related with the mean free path of a nucleon inside a nucleus and has a constant value for nucleus-nucleus collisions.
In the present calculation, the midrapidty pseudorapidity density for symmetric nuclei like $Au$-$Au$, $Cu$-$Cu$, $Pb$-$Pb$ has been calculated using the centrality division for quark-nucleus inelastic cross-section, ($\sigma_{qA}^{in}$) as ginen in refs.~\cite{Ashwini_2:2013,Ashwini:2013,Chaturvedi:2016,Chaturvedi_2017}, and for asymmetric nuclei like ($d$-$Au$ and $p$-$Pb$), we have used the centrality division for quark-nucleus inelastic cross-section, ($\sigma_{qA}^{in}$) as obtained in the refs. ~\cite{Ashwini:2013,Chaturvedi:2016}. Having these values of $\sigma_{AB}^{in}$, we have obtained the value of $\left(\frac{dn_{ch}}{d\eta}\right)^{AB}_{\eta=0}$ using Eq. 12.  

\begin{figure}
\includegraphics[scale=0.52]{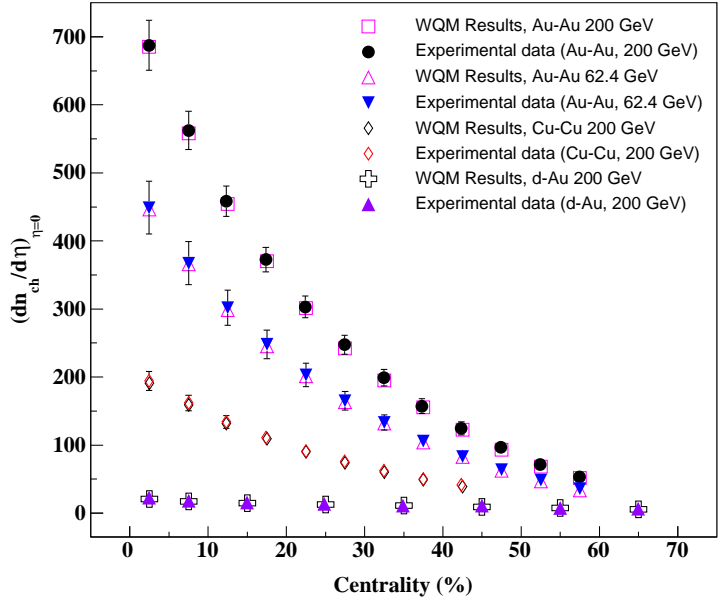}
\caption{(Color online) Variation of the pseudo-rapidity density at midrapidity of charged hadrons with respect to centrality for different colliding systems. The experimental data used here are taken from Ref. ~\cite{Adler:2004zn,Adare_2016_PRC_1}.}
\label{fig1}
\end{figure}

\begin{figure}
\includegraphics[scale=0.52]{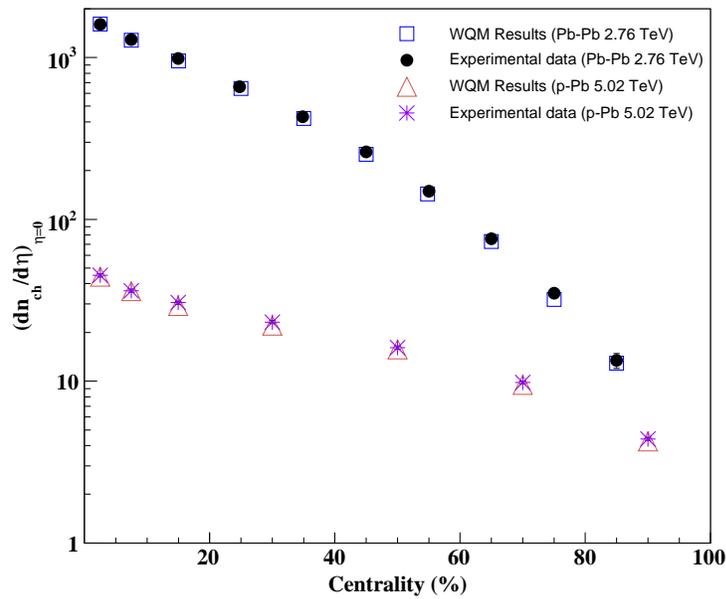}
\caption{(Color online) Variation of the pseudo-rapidity density at midrapidity of charged hadrons for $Pb$-$Pb$ ~\cite{Adam_2016_PRC} and $p$-$Pb$ ~\cite{Adam_2016} collisions with respect to centrality at LHC energies.}
\label{fig2}
\end{figure}

\begin{figure}
\includegraphics[scale=0.52]{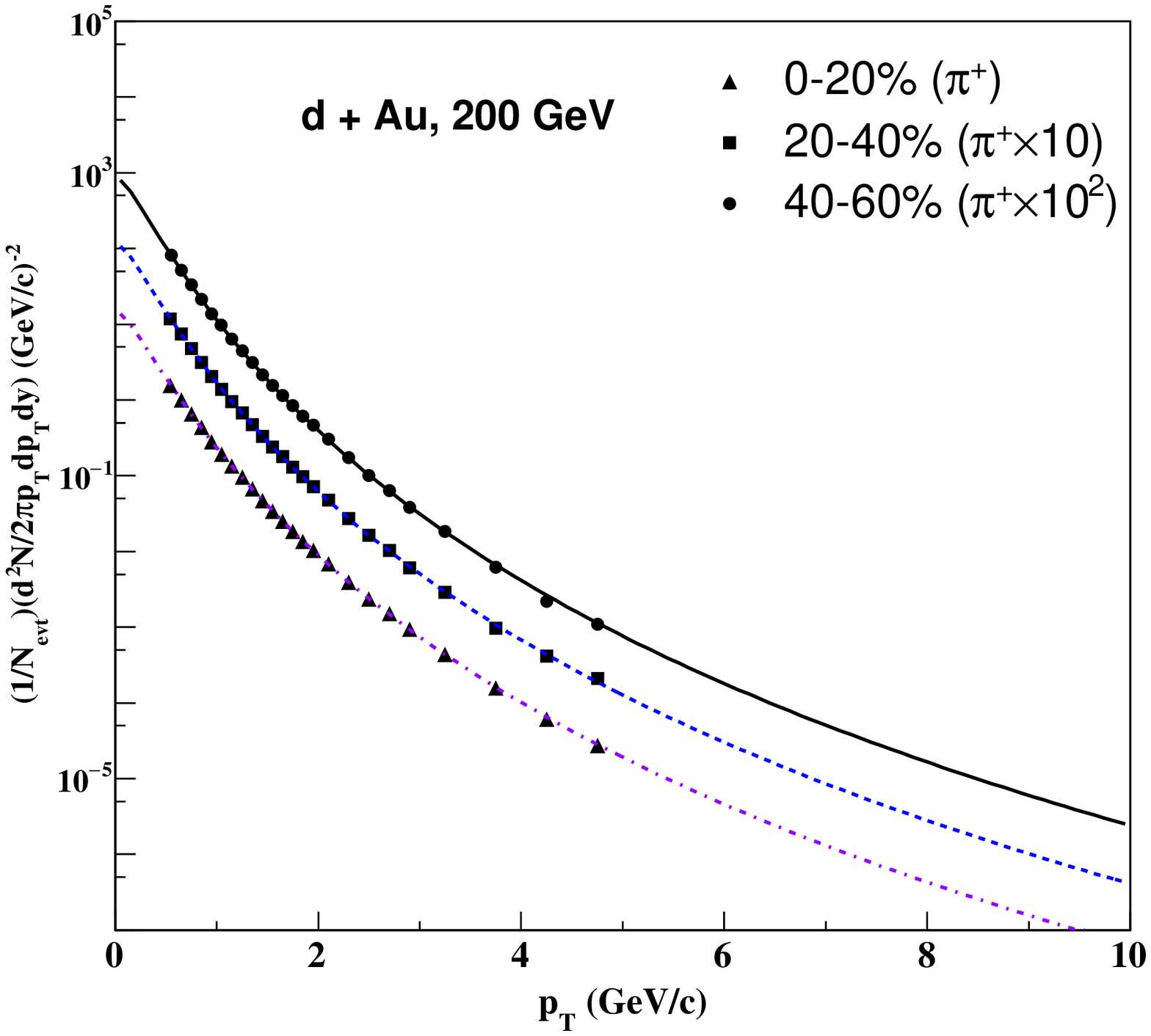}
\caption{(Color online) The invariant yields of charged positive pions as a function of $p_{T}$ for $d$-$Au$ collision at 200 GeV ~\cite{Adare_2013}. The different curves are the result of fitted Tsallis distribution, Eq. 7.}
\label{fig3}
\end{figure}

\begin{figure}
\includegraphics[scale=0.52]{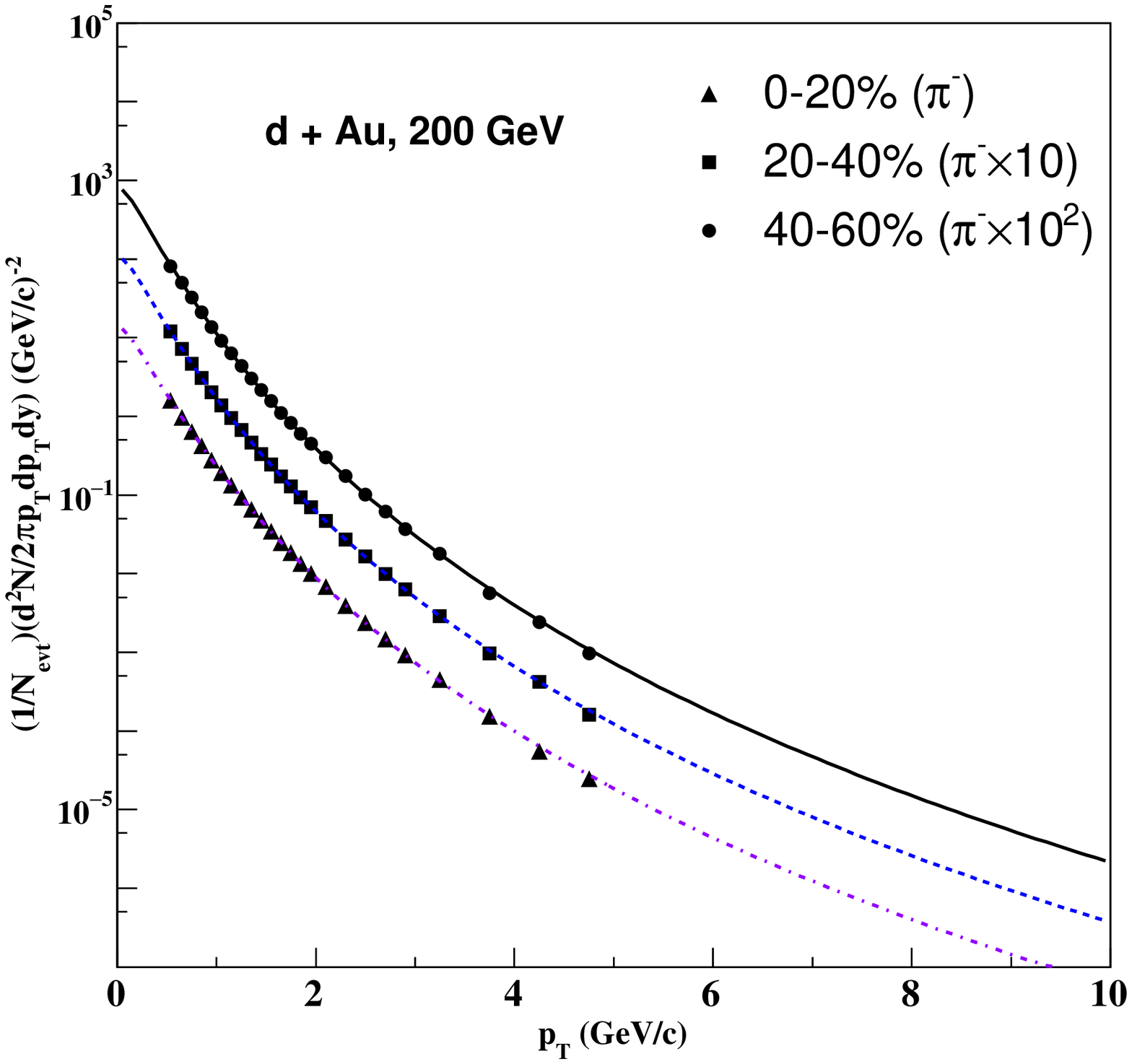}
\caption{(Color online) The invariant yield of charged negative pions as a function of $p_{T}$ for $d$-$Au$ collision at 200 GeV ~\cite{Adare_2013}. The different curves are the result of fitted Tsallis distribution, Eq. 7.}
\label{fig4}
\end{figure}

\begin{figure}
\includegraphics[scale=0.52]{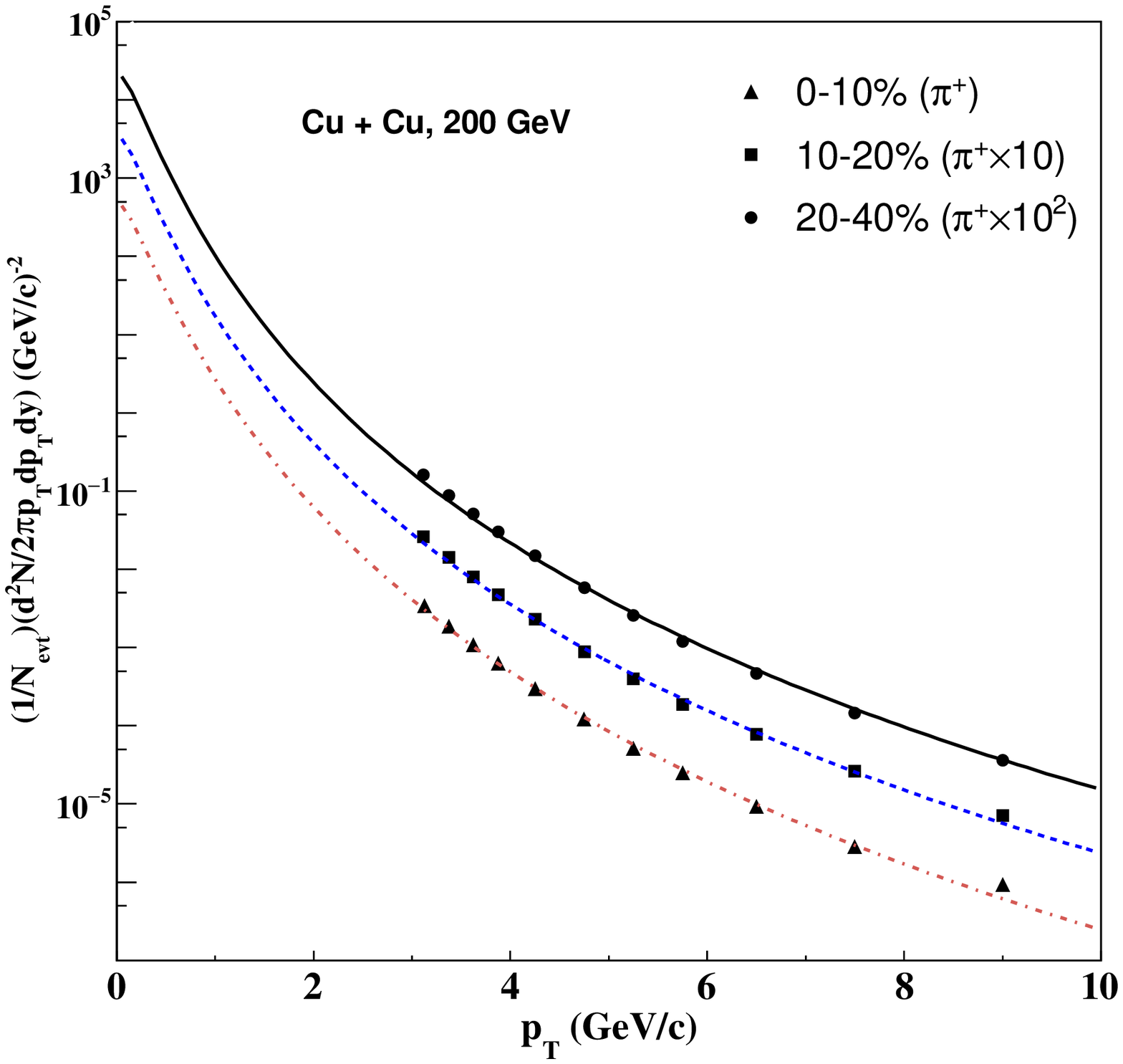}
\caption{(Color online) The invariant yield of charged positive pions as a function of $p_{T}$ for $Cu$-$Cu$ collision at 200 GeV ~\cite{Abelev_2010}. The different curves are the result of fitted Tsallis distribution, Eq. 7.}
\label{fig5}
\end{figure}

\begin{figure}
\includegraphics[scale=0.52]{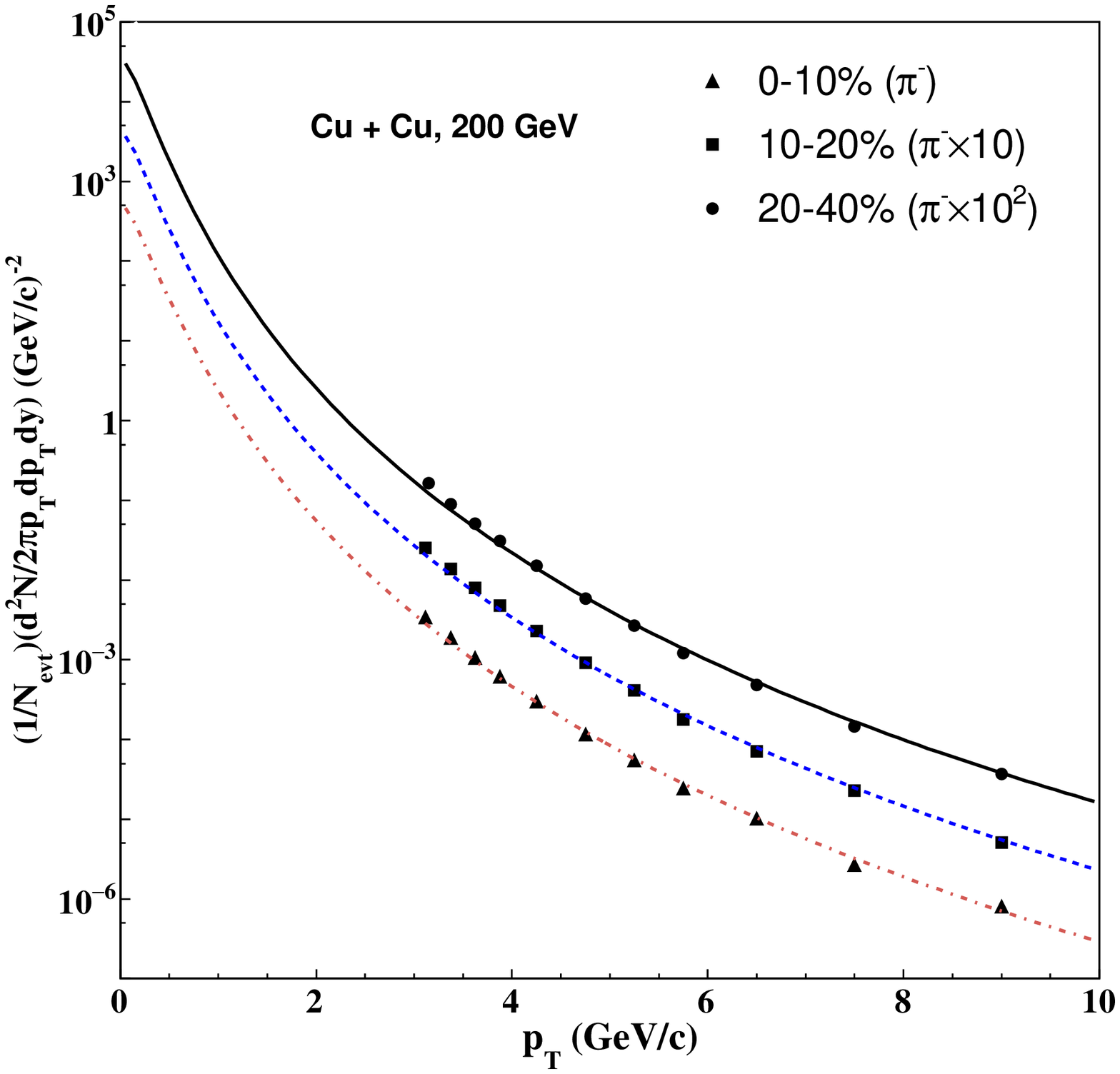}
\caption{(Color online) The invariant yield of charged negative pions as a function of $p_{T}$ for $Cu$-$Cu$ collision at 200 GeV ~\cite{Abelev_2010}. The different curves are the result of fitted Tsallis distribution, Eq. 7.}
\label{fig6}
\end{figure}

\begin{figure}
\includegraphics[scale=0.52]{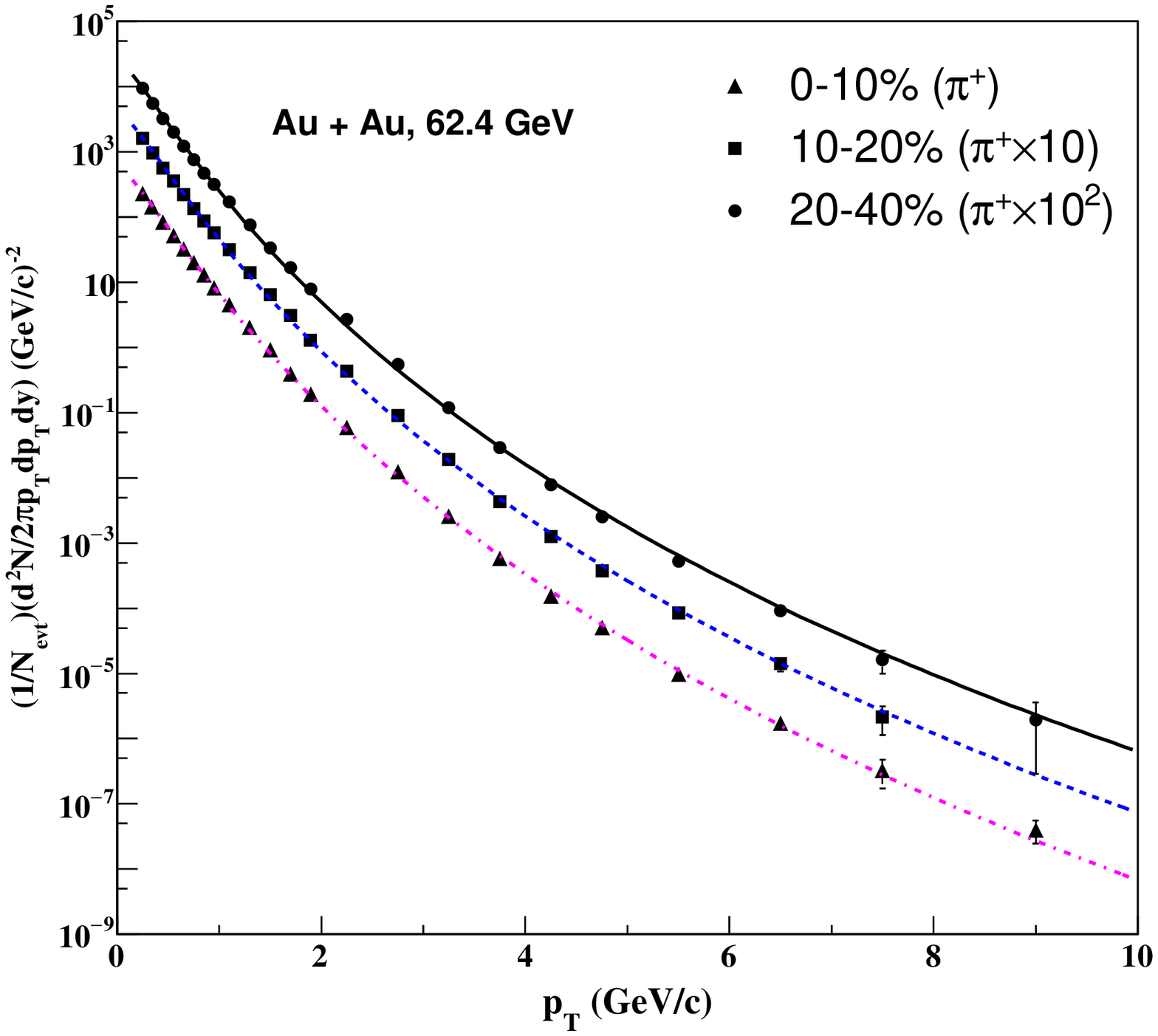}
\caption{(Color online) The invariant yield of charged positive pions as a function of $p_{T}$ for $Au$-$Au$ collision at 62.4 GeV ~\cite{Abelev_2007}. The different curves are the result of fitted Tsallis distribution, Eq. 7.}
\label{fig7}
\end{figure}

\begin{figure}
\includegraphics[scale=0.52]{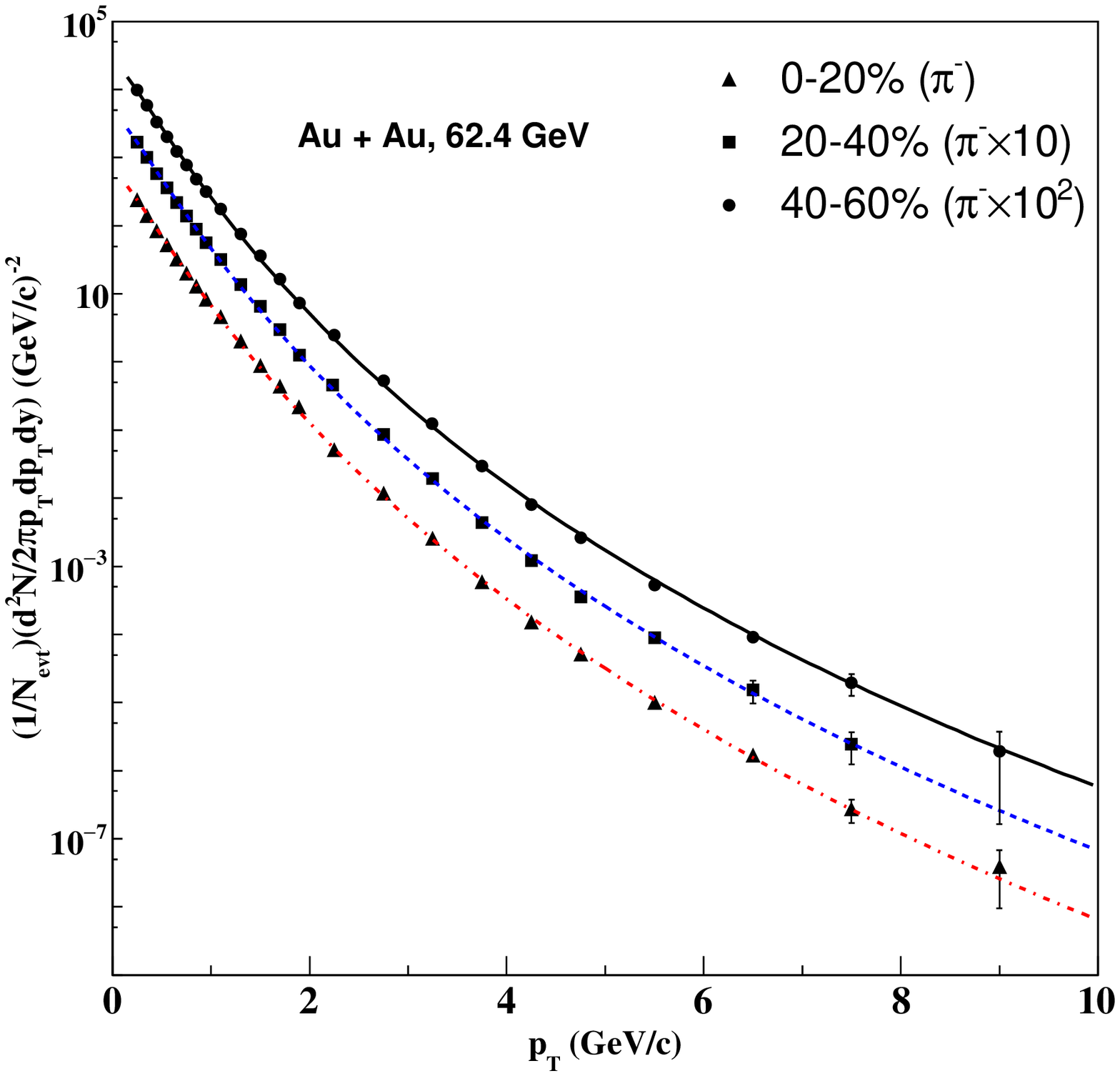}
\caption{(Color online) The invariant yield of charged negative pions as a function of $p_{T}$ for $Au$-$Au$ collision at 62.4 GeV ~\cite{Abelev_2007}. The different curves are the result of fitted Tsallis distribution, Eq. 7.}
\label{fig8}
\end{figure}

\begin{figure}
\includegraphics[scale=0.52]{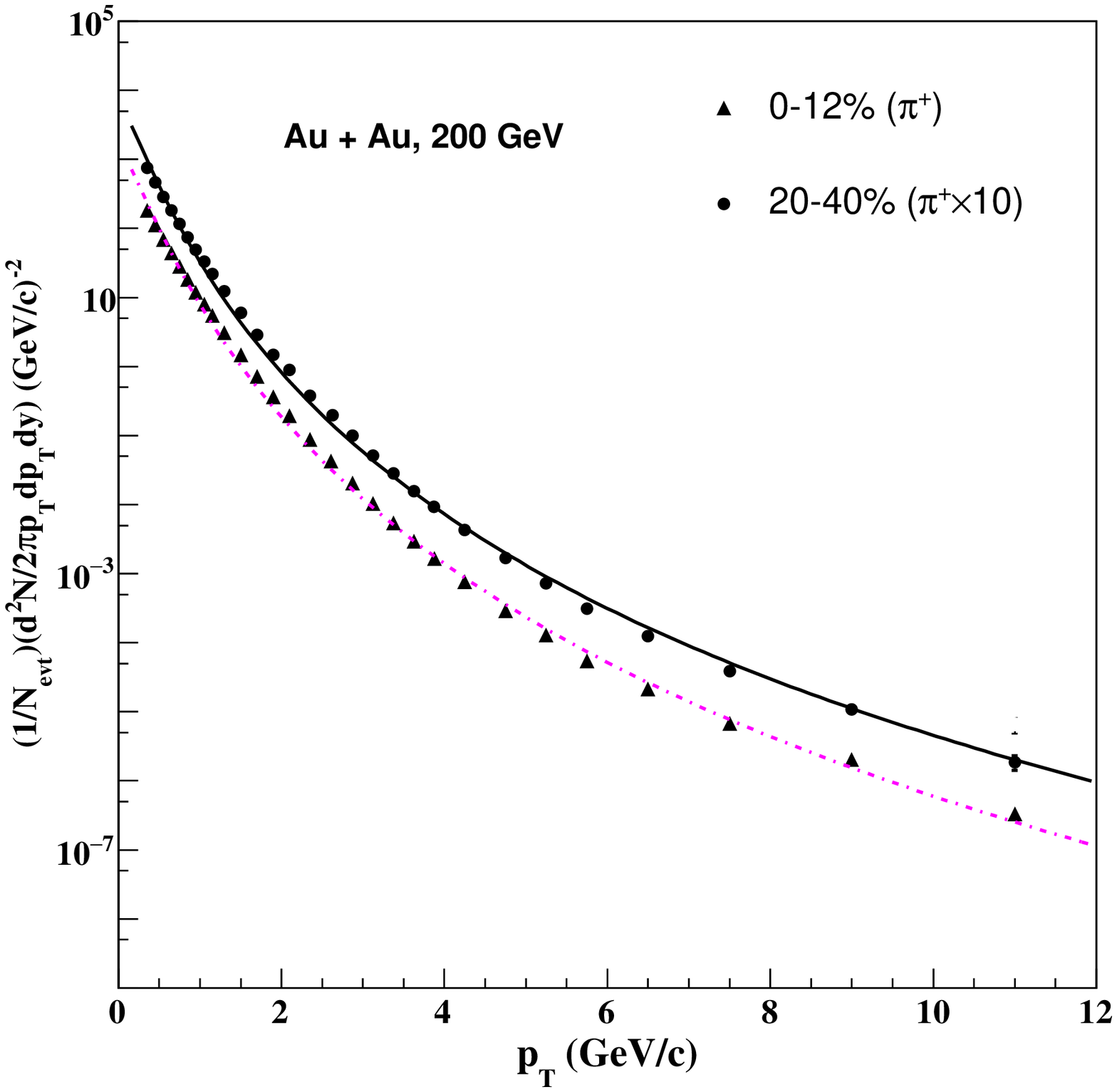}
\caption{(Color online) The invariant yield of charged positive pions as a function of $p_{T}$ for $Au$-$Au$ collision at 200 GeV ~\cite{Abelev_2007}. The different curves are the result of fitted Tsallis distribution, Eq. 7.}
\label{fig9}
\end{figure}

\begin{figure}
\includegraphics[scale=0.52]{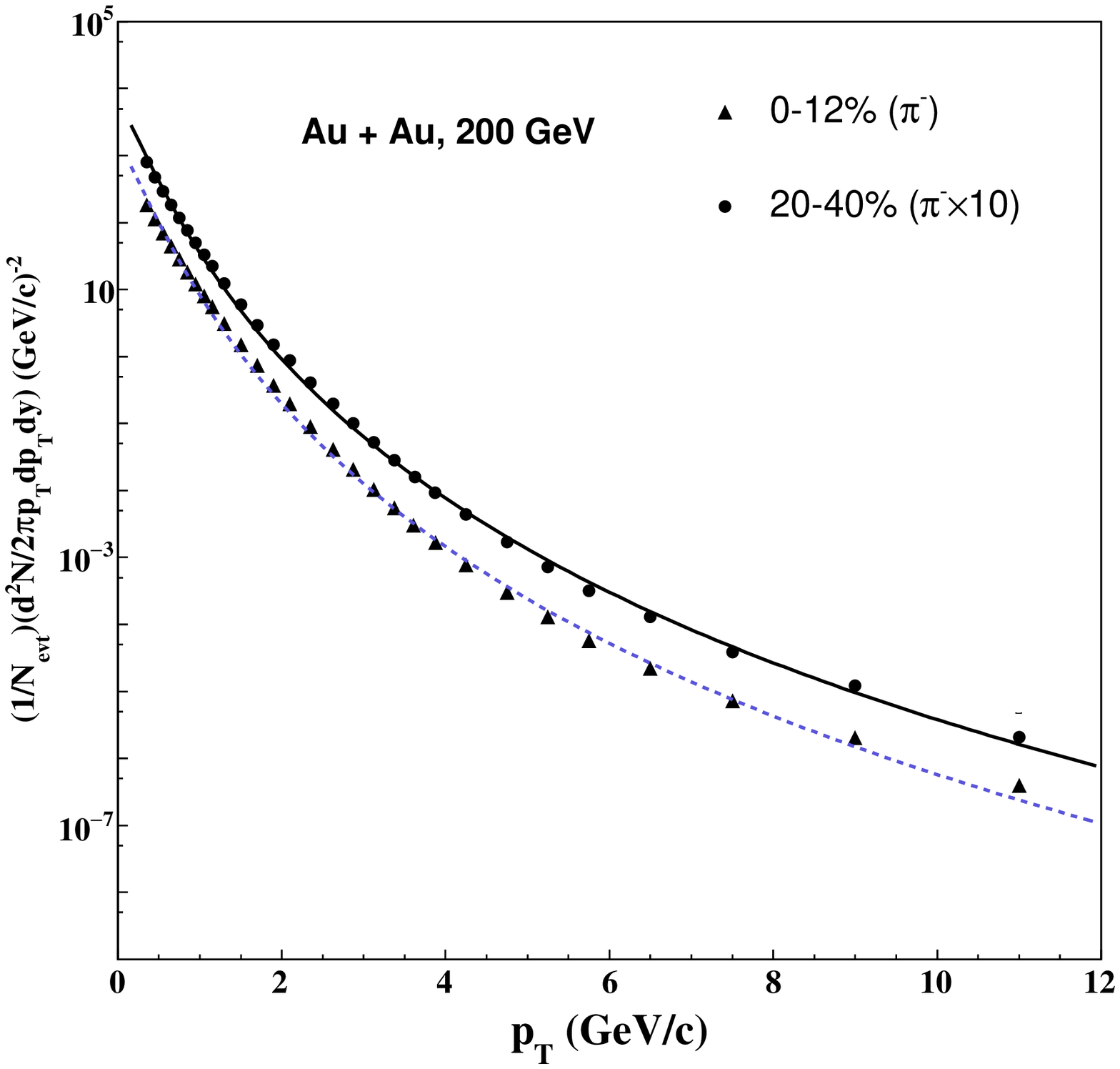}
\caption{(Color online) The invariant yield of charged negative pions as a function of $p_{T}$ for $Au$-$Au$ collision at 200 GeV ~\cite{Abelev_2007}. The different curves are the result of fitted Tsallis distribution, Eq. 7.}
\label{fig10}
\end{figure}

\begin{figure}
\includegraphics[scale=0.52]{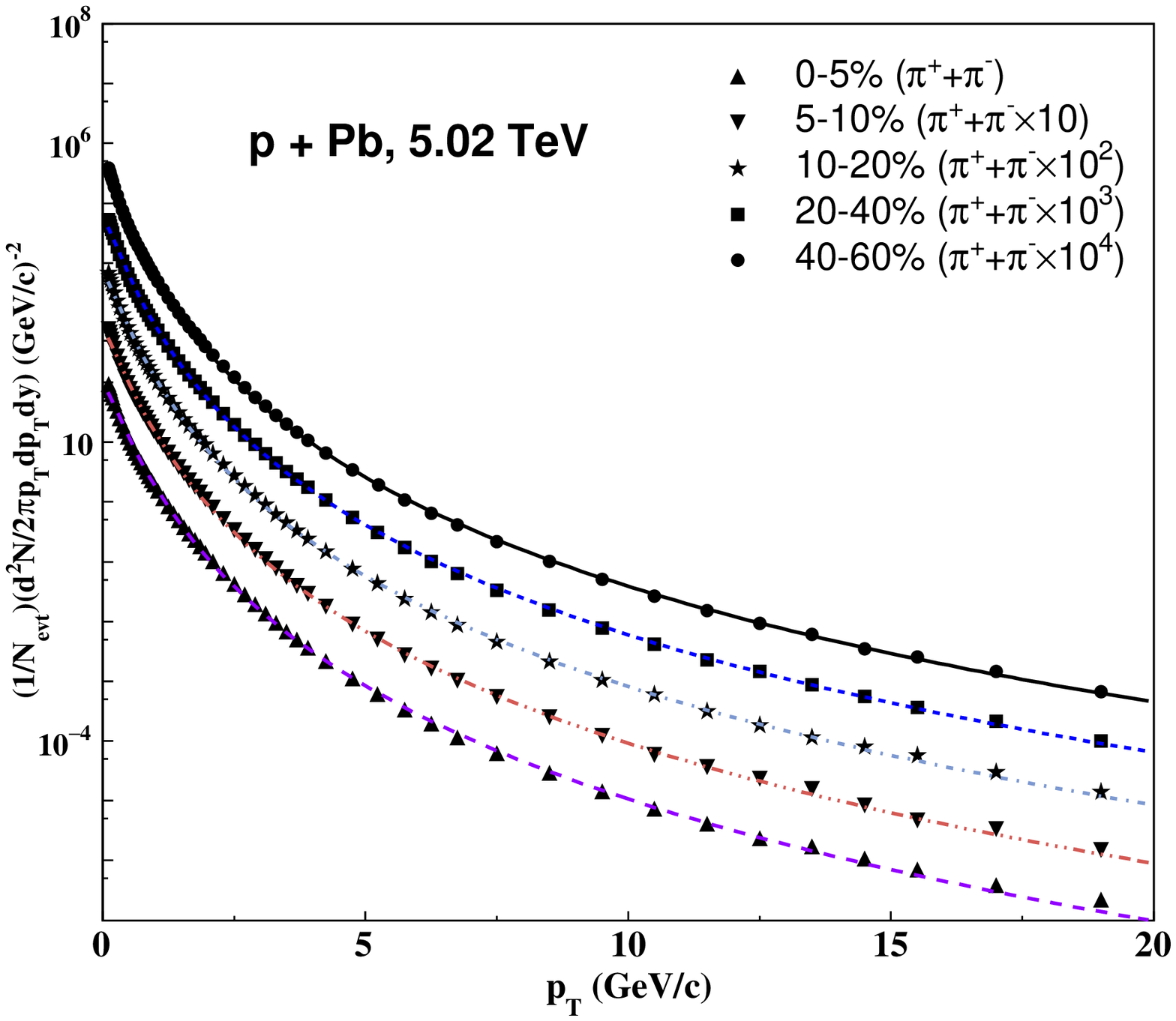}
\caption{(Color online) The invariant yield of charged pions as a function of $p_{T}$ for $p$-$Pb$ collision at 5.02 TeV ~\cite{Adam_2016}. The different curves are the result of fitted Tsallis distribution, Eq. 7.}
\label{fig11}
\end{figure}

\begin{figure}
\includegraphics[scale=0.52]{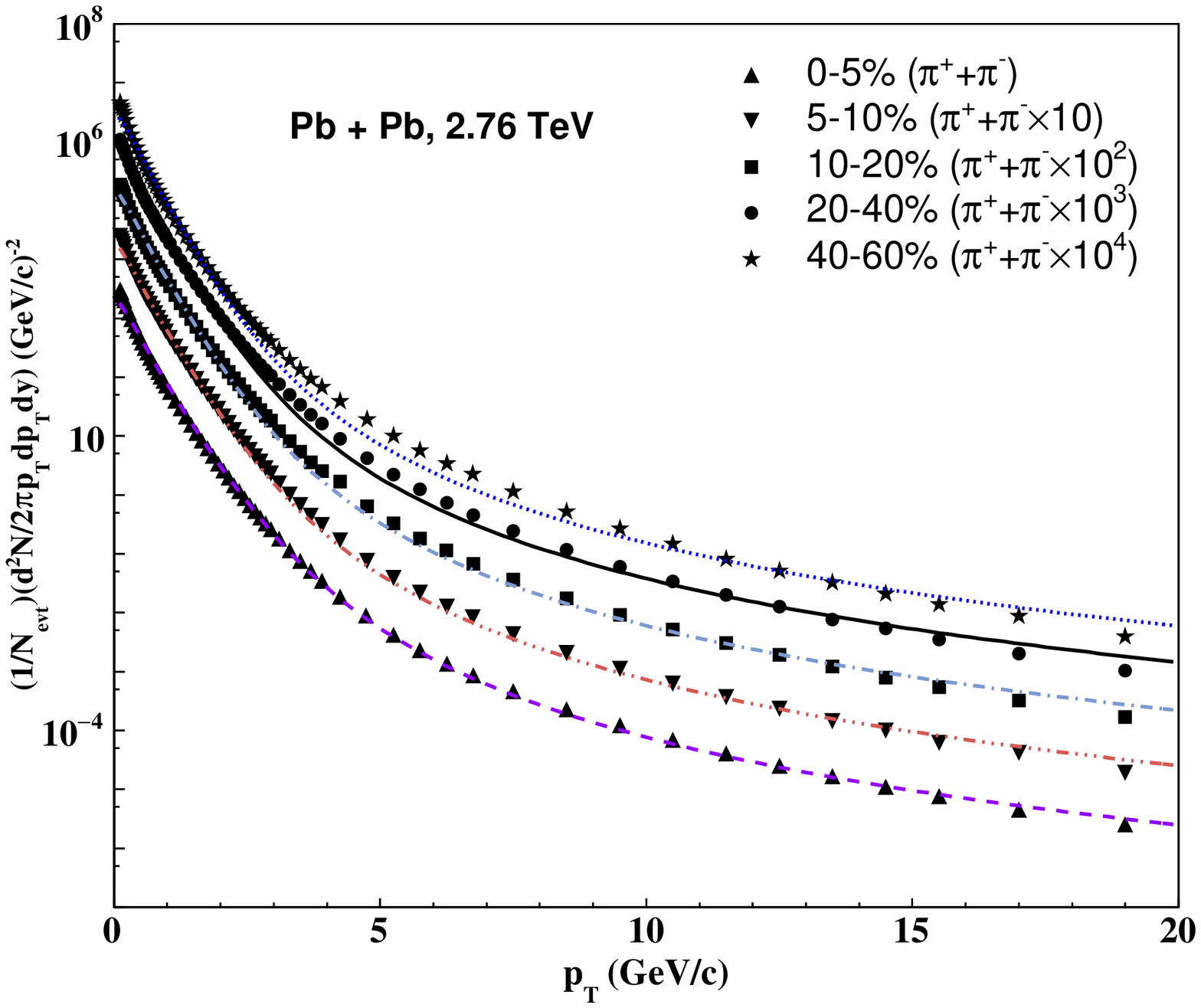}
\caption{(Color online) The invariant yield of charged pions as a function of $p_{T}$ for $Pb$-$Pb$ collision at 2.76 TeV ~\cite{Adam_2016_PRC}. The different curves are the result of fitted Tsallis distribution, Eq. 8.}
\label{fig12}
\end{figure}

\begin{figure}
\includegraphics[scale=0.52]{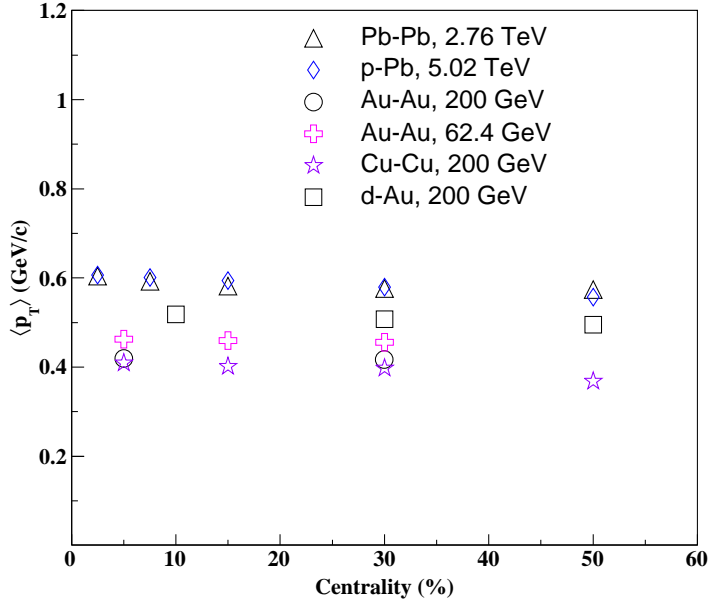}
\caption{(Color online) Variation of the midrapidity average transverse momentum, $\left \langle p_{T} \right \rangle_{y=0}$, obtained using Eq. 11, as a function of centrality for different colliding systems.}
\label{fig13}
\end{figure}

\begin{figure}
\includegraphics[scale=0.52]{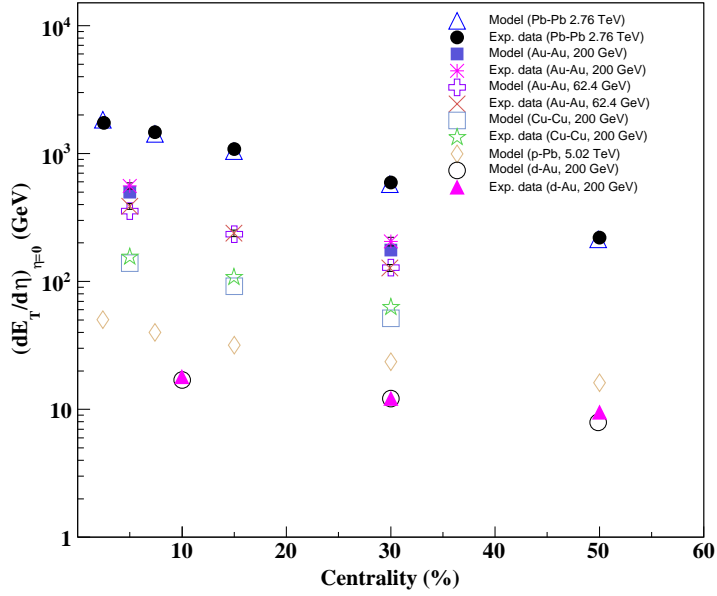}
\caption{(Color online) Variation of the midrapidity transverse energy density, $(dE_{T}/d\eta)_{\eta=0}$ as a function of centrality for different colliding systems. The different experimental data are taken from Ref. ~\cite{Adare_2016_PRC_1}.}
\label{fig14}
\end{figure}

\begin{figure}
\includegraphics[scale=0.52]{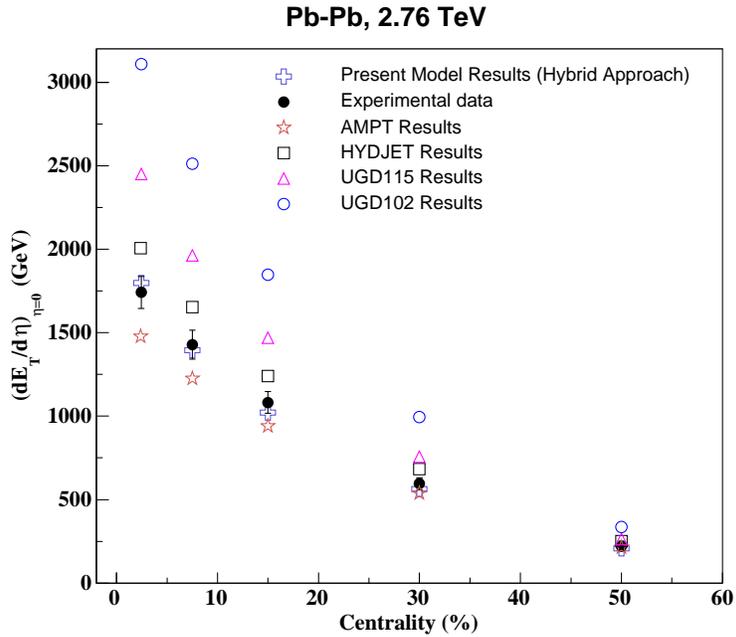}
\caption{(Color online) Comparison of the midrapidity transverse energy density, $(dE_{T}/d\eta)_{\eta=0}$ obtained from present calculation as a function of centrality for $Pb$-$Pb$ with different types of theoretical and phenomenological models ~\cite{Lin_2001,Lokhtin_2006,Albacte_2013}. The experimental data is taken from Ref. ~\cite{Adam_2016_PRC}.}
\label{fig15}
\end{figure}

\begin{figure}
\includegraphics[scale=0.52]{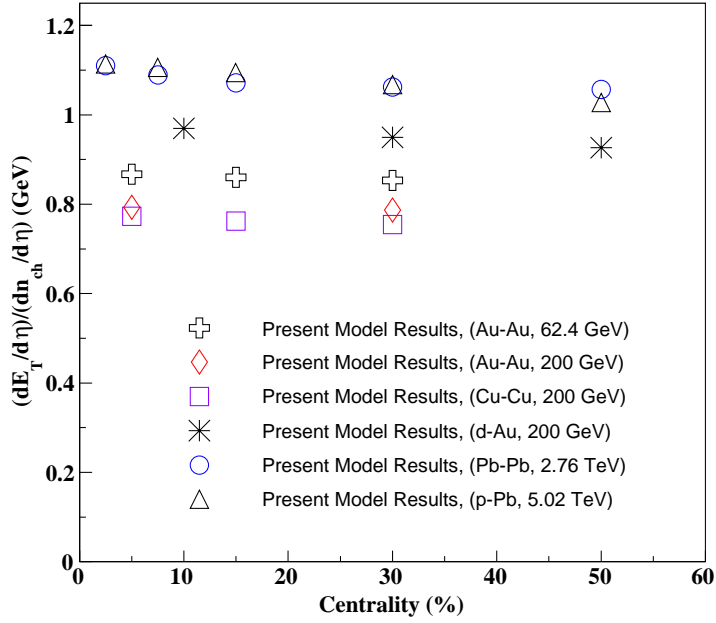}
\caption{(Color online) Variation of the ratio of transverse energy to total mean multiplicity ($E_{T} /N_{ch}$ $\equiv(dE_{T}/d\eta)/(dn_{ch}/d\eta)$), obtained by the Eq. 16 as a function of centrality for different colliding systems.}
\label{fig16}
\end{figure}

\begin{figure}
\includegraphics[scale=0.52]{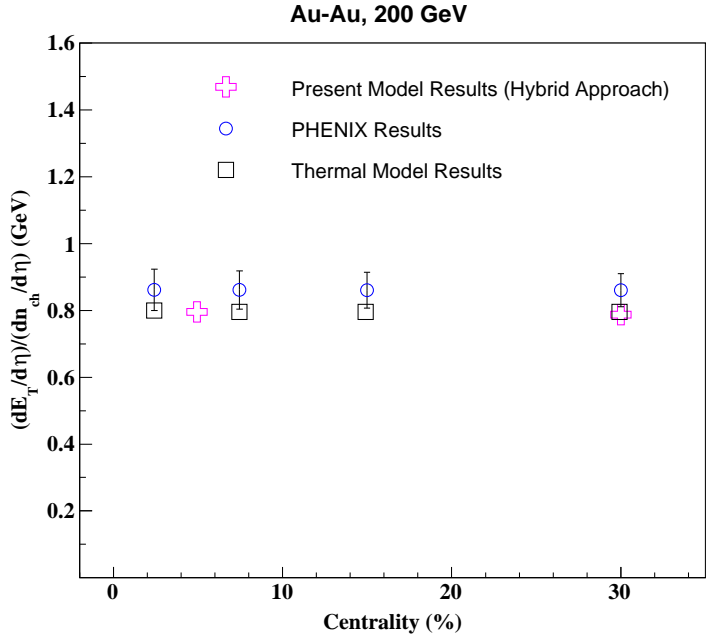}
\caption{(Color online) Comparison of $E_{T} /N_{ch}$, obtained by Eq. 16, with PHENIX experimental data and Thermal model ~\cite{skt2} as a function of centrality for $Au$-$Au$ at 200 GeV ~\cite{Adare_2016_PRC_1}.}
\label{fig17}
\end{figure}

\begin{figure}
\includegraphics[scale=0.52]{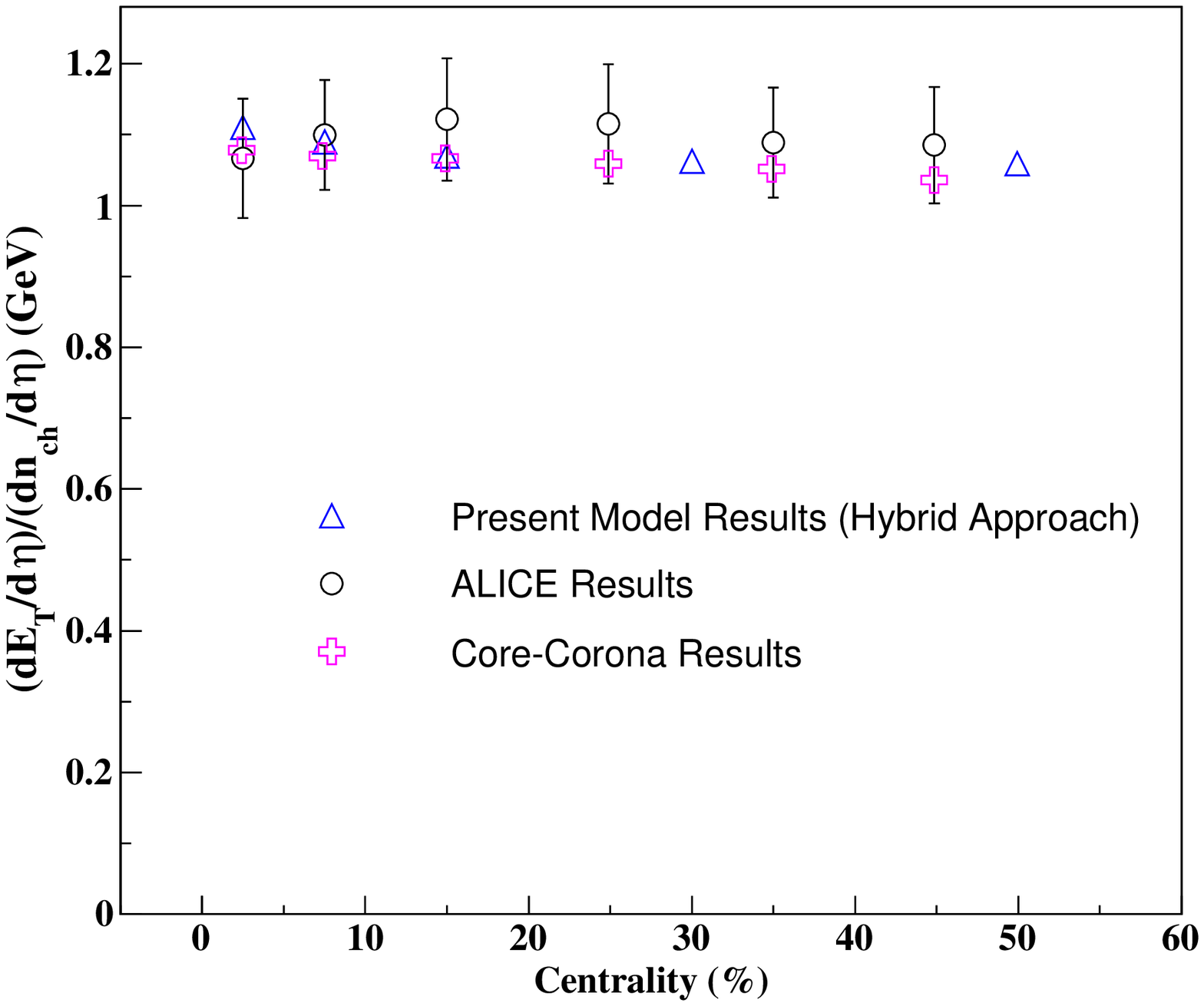}
\caption{(Color online) Comparison of $E_{T} /N_{ch}$, obtained by Eq. 16, with ALICE experiment and Core-Corona results ~\cite{Petrovici_17} as a function of centrality for $Pb$-$Pb$ at 2.76 TeV ~\cite{Adam_2016_PRC}.}
\label{fig18}
\end{figure}

\subsection{Transverse Energy Density Distribution}
The transverse energy density distribution of charged hadrons ~\cite{Bjorken_1983} can be calculated using the pseudorapidity distribution of WQM and $\left\langle p_{T} \right \rangle$ values obtained using Tsallis parameters can be given as
\begin{equation}
dE_{T}/d\eta \cong \frac{3}{2} \sqrt{{\left\langle p_{T} \right \rangle}^2+m_{\pi}^2} (dn_{ch}/d\eta),
\end{equation}
where $\langle p_{T}\rangle$ is the average transverse momentum of the produced charged particles and $m_{\pi}$ is the mass of pion.

\section{Results and Discussions}
\subsection{Pseudorapidity Density and Transverse Momentum Distributions}
In this section we have shown the results obtained from WQM regarding $dn_{ch}/d\eta$ at mid-
rapidity and the results regarding transverse momentum distributions using Tsallis distri-
bution function for various type of collisions. In Fig. ~\ref{fig1}, we have plotted the
variation of $dn_{ch}/d\eta$ at mid-rapidity with respect to centrality for various colliding species
e.g., $Au$-$Au$, $Cu$-$Cu$, $d$-$Au$ etc. at different collision energies. We have compared our model
results with the corresponding experimental data and found suitable match. From this
graph one can observe that the effect of change in energy (see $Au$-$Au$ at $62.4$ and $200$ GeV)
on the charged hadron multiplicity is not quite substantial in comparison to the effect of
change in colliding species at the same energy (see $Au$-$Au$,$Cu$-$Cu$ and $d$-$Au$ at $200$ GeV).
Fig.~\ref{fig2} demonstrates the variation of $dn_{ch}/d\eta$ at mid-rapidity with respect to centrality for $Pb$-$Pb$ collisions at $2.76$ and for $p$-$Pb$ at $5.02$ TeV. Our model results satisfy the
experimental data quite well. 
In Fig.~\ref{fig3}, we have plotted the experimental data for normalized invariant yield of positively charged pions produced in $d$-$Au$ collisions at $200$ GeV with respect to $p_{T}$. We
have shown these data for three centrality class starting from central ($0-20\%$) to peripheral
($40-60\%$) collisions. Further, we have used the Tsallis distribution function to fit this invariant yield. The value of parameters obtained is shown in table I. We have also fitted
the invariant yield of negatively charged pions produced in $d$-$Au$ collisions at $200$ GeV in
Fig.~\ref{fig4} and the fitting parameters for Tsallis distribution is mentioned in table I.

Similarly Fig.~\ref{fig5} and ~\ref{fig6} show the fitting of Tsallis distribution through experimental data
of positively and negatively charged pions in $Cu$-$Cu$ collisions at $200$ GeV. Here again we
have fitted the invariant yield distribution in three centrality class i.e., $0-10\%$, $10-20\%$
and $20-40\%$. Fig.~\ref{fig7} and ~\ref{fig8} show the fitting of Tsallis distribution through experimental data of positively and negatively charged pions in $Au$-$Au$ collisions at $62.4$ GeV. Here again we have fitted the invariant yield distribution in three centrality class i.e., $0-20\%$, $20-40\%$
and $40-60\%$. In Fig.~\ref{fig9} and ~\ref{fig10}, we have demonstrated the Tsallis distribution fit through experimental data for normalized $p_{T}$-distribution of $\pi^{+}$ and $\pi^{-}$ produced in $Au$-$Au$ collisions at $200$ GeV in two different centrality classes.  

Fig.~\ref{fig11}, presents the Tsallis fit through experimental data for normalized $p_{T}$-distribution of $\pi=\pi^{+}+\pi^{-}$ produced in $p$-$Pb$ collisions at $5.02$ TeV. We have shown the results for five different centrality classes from most central ($0-5\%$) to most peripheral ($40-60\%$). Further in Fig.~\ref{fig12}, we have shown the Tsallis fit with a modified Tsallis distribution (as given by Eq. (8)) for normalized $p_{T}$-distribution of $\pi=\pi^{+}+\pi^{-}$ produced in $Pb$-$Pb$ collisions at $2.76$ TeV. We have again shown the results for five different centrality classes from most central ($0-5\%$) to most peripheral ($40-60\%$).  

As we see by comparing eq. 7 and eq. 3, there is a clear dependence of the parameter $q$ on $n$; both are related by the relation $n=1/q-1$. If we analyse the different values of $n$ from the table I at different collisional energy and for different colliding systems, we can understand the variation of parameter $n$ or $q$ with these control parameters. For $Au$-$Au$ at $\sqrt{s_{NN}} = 200$ GeV the value of $n$ is approximately seen to be $10$, so we can calculate the q value which comes out as $11/10=1.11$. Similarly for $d$-$Au$ at $\sqrt{s_{NN}}= 200$ GeV, the value of n is again around 10 (see table I) and again the q value comes out as $1.11$. For $Cu$-$Cu$ at $200$ GeV, we again find same value of $q$. From all these observations we find that $q$ for different colliding system at $200$ GeV have same value and thus suggest that parameter $q$ is independent to the colliding systems. Now if we move to $Au$-$Au$ at $\sqrt{s_{NN}}$ = 62.4 GeV then $n$ takes a value of $1.058$ which gives $q=1.05$ and for $p$-$Pb$ system at $5.02$ TeV the value of $q$ goes to 1.15. From these measurement we see the value of $q$ increase from $1.058$ to $1.15$ as we increase the collision energy from $62.4$ GeV to $5.02$ TeV. This observation suggest that the values of parameter $q$ is not equal to unity. The increase in $q$ value from unity causes the transverse momentum distribution to deviate from exponential distribution function to the power law and further this deviation increases with increase in collision energy. In Tsallis parametrization, parameter $A$ is equal to $Cdn/dy$, where $C$ is constant and $dn/dy$ is rapidity density. The dependence of $A$ can help us to quantify the multiplicity in basic hadron-hadron collisions at those energies. The parameter $A$ (as listed in table I and 2) for $Au$-$Au$ takes the value from $1210$ to $566$ at $200$ GeV and $537.3$ to $224.7$ at $62.4$ GeV with respect to centrality. For $d$-$Au$ at $200$ GeV, it goes from $14$ to $8$ as we move towards peripheral collisions from central collisions. These observations suggest that $A$ depends on the colliding system as well as on colliding energy. Coming towards the third parameter $T$ which have different values for different colliding systems as shown in the table I and table II. For a particular system as we go from central to peripheral collisions the value of $T$ decrease from a maximum to minimum value; which shows that the fireball formed in the nuclear collisions becomes less denser as we move from central to peripheral collision. This behaviour is pertinent for all the colliding systems. For $Pb$-$Pb$ collision at $\sqrt{s_{NN}} = 2.76$ TeV we have used four parameters $A,~T,~b$ and $c$. The values of $A$ and $T$ coming similar as observed in previous collisions and they have similar physical explanation, for other two parameter $a$ and $b$ which serve qualitatively as the parameter $n$ show an energy dependence.

\subsection{Average $p_{T}$ and Transverse Energy Density Distributions}
In this section we have shown the results regarding the thermal average of transverse momentum for the produced charged hadrons in various type of collisions and different collision energies. $\langle p_{T}\rangle$ is calculated by using Eq. (11). In Fig.~\ref{fig13}, we have shown the $\langle p_{T}\rangle$ for various collision types. Here we find that the $\langle p_{T}\rangle$ varies between $0.4$ to $0.6$ with change in collisions energy and colliding species. In assymetric collisions like $d$-$Au$ and $p$-$Pb$, the system created is of very small size and thus in these collisions the finite size effects are dominant. This actually make the thermal averages to vary in different way than in bigger system which is created during $Cu$-$Cu$, $Au$-$Au$ and $Pb$-$Pb$ collisions.
In Fig.~\ref{fig14}, we have presented the variation of transverse energy density distribution ($dE_{T}/d\eta$) at midrapidity with respect to centrality. We have calculated $dE_{T}/d\eta$ at $\eta =0$ using $\langle p_{T}\rangle$ as shown in Fig.~\ref{fig13} and $dn_{ch}/d\eta$ as shown in Fig.~\ref{fig1} and ~\ref{fig2}. We have shown $dE_{T}/d\eta$ for various colliding species at different collision energies. Here, one can see that $d$-$Au$ collisions produced the least values of transverse energy densities and $Pb$-$Pb$ collisions produced the highest values of $dE_{T}/d\eta$ among the colliding species considered in the present calculation. We have compared our hybrid model results with the data obtained from various collision experiments. We observe that our model results suitably matches with the corresponding experimental data. One important observation is that the transverse energy density in central $p$-$Pb$ collisions at $5.02$ TeV is nealy comparable to the transverse energy density produced in $Cu$-$Cu$ collisions at $200$ GeV. In Fig.~\ref{fig15}, we have presented a comparison of various model results with our hybrid approach regarding  $dE_{T}/d\eta$ at $\eta =0$ in $Pb$-$Pb$ collisions at $2.76$ TeV. We have also plotted the experimental data for comparison. From this figure, one can see that the hybrid approach describes the data most suitably. AMPT (a multi-phase transport model) and HYDJET (hydro plus jet) model also satisfy the experimental data except at central collisions. However, UGD 115 and UGD 102 model results clearly overestimate $dE_{T}/d\eta$ at $\eta =0$.

\subsection{$E_{T}/N_{ch}$ : A Freezeout Criteria}
In this last part of the results and discussion section, we have plotted the ratio $(dE_{T}/d\eta)/(dn_{ch}/d\eta)$ which is equivalent to $E_{T}/N_{ch}$ and is considered as a freezeout condition in heavy-ion collision experiments. In Fig.~\ref{fig16}, we have presented this ratio with respect to centrality for various colliding species at different collision energies. From figure, it is clear that for smaller energies, this ratio is below $1$ and for higher energies, $E/N$ is larger than $1$. Varying the collision energy from $200$ GeV to $5.02$ TeV, this ratio only varies in the range $0.8$ to $1.1$. Thus our study suggest that $E/N$ varies but still it is a suitable freezeout criteria. Based on our present hybrid approach, we suggest that $E_{T}/N_{ch}\approx 0.95\pm 0.15$ can act as a freezeout criteria for bulk studies regarding non-strange charged hadrons in heavy-ion collisions. However, from this study we can not say whether $E_{T}/N_{ch}\approx 0.95\pm 0.15$ is a robust freezeout criteria to study the strange and/or charm physics in heavy-ion collisions. This physical interpretation of this criteria is that the chemical freezeout during the evolution of medium occurs when the transverse energy per particle (or charged hadron) comes down to $0.95\pm 0.15$. If the transverse energy per particle is lagrer than this value then the inelastic collisions are still there among the hadrons. Another important observation from this plot is that the condition $E_{T}/N_{ch}=0.95\pm 0.15$ is satisfied by charged hadrons produced even in the $d-Au$ and $p-Pb$ collisions. Thus this criteria is robust even for small systems. In Fig.~\ref{fig17}, we have shown our model results of $E_{T}/N_{ch}$ obtained in $Au$-$Au$ collision at $200$ GeV and compared it with the PHENIX experimental data~\cite{Adare_2016_PRC_1} and a thermal model approach~\cite{skt2}. We have done our calculation only for two centrality classes. The present hybrid approach and the thermal model both satisfy the experimental data. The value of $E_{T}/N_{ch}$ is equal to $0.8$ at this collision energy. Similarly in Fig.~\ref{fig18}, we have compared our hybrid model results for $E_{T}/N_{ch}$ in $Pb$-$Pb$ collision at $2.76$ TeV with the experimental data from ALICE collaboration~\cite{Adam_2016_PRC} and the corresponding results from core-corona model~\cite{Petrovici_17}. Here we want to remind that core-corona model is observed as a precise model to discuss the average thermal momentum in heavy-ion collisions. We found that our hybrid approach properly satisfy the core-corona model results along with ALICE data. The value of $E_{T}/N_{ch}$ at this $2.76$ TeV energy is almost equal to $1.1$. Thus we can say that our hybrid approach describes the $E_{T}/N_{ch}$ result at $200$ GeV and at $2.76$ TeV, simultaneously.

In summary, we have constructed a hybrid model in which the pseudorapidity distribution is derived from wounded quark model (WQM) and transverse momentum distribution is obtained from Tsallis statistical model. We have first calculated the pseudorapidity distribution at mid-rapidity using WQM with respect to centrality for various colliding species at different collision energies. Further, we have fitted the transverse momentum distributions for different collisions and obtained a fitting parametrization along with its parameters in these various collisions and use this Tsallis parametrization to calculate the average transverse momentum $\langle p_{T}\rangle$. After that we have calculated and plotted the transverse energy density distributions with respect to centrality for charged hadrons in various type of collisions. At last we have plotted the ratio $(dE_{T}/d\eta)/(dn_{ch}/d\eta)~=~E_{T}/N_{ch}$ and is considered as a freezeout criteria for charged hadrons in heavy-ion collision experiments. We have compared this ratio as obtained in our hybrid model and compare them with the experimental data as well as with the results from other phenomenological models. We observed that the condition on transverse energy per charged hadron i.e., $E_{T}/N_{ch}=0.95\pm 0.15$ can act as a robust freezeout criteria for charged hadron production in high-energy nuclear collisions for small as well as for large systems.

\section{Acknowledgments}
\noindent 
OSKC is grateful to Council of Scientific and Industrial Research (CSIR), New Delhi for providing a research grant. PKS acknowledges IIT Ropar, India for providing an institute postdoctoral research grant. 


\end{document}